# Variable Potentials for Thermalized Light and Coupled Condensates

David Dung, Christian Kurtscheid, Tobias Damm, Julian Schmitt, Frank Vewinger, Martin Weitz & Jan Klärs[†]

*Institut für Angewandte Physik, Universität Bonn, Wegelerstr. 8, 53115 Bonn, Germany*

**For over a decade, cold atoms in lattice potentials have been an attractive platform to simulate phenomena known from solid state theory, as the Mott-insulator transition[1]. In contrast, the field of photonics usually deals with non-equilibrium physics[2-5]. Recent advances towards photonic simulators of solid state equilibrium effects include polariton double-site and lattice experiments[6-10], as well as the demonstration of a photon condensate in a dye-filled microcavity[11,12]. Here we demonstrate a technique to create variable micropotentials for light using thermo-optic imprinting within an ultrahigh-reflectivity mirror microcavity filled with a dye-polymer solution that is compatible with photon gas thermalization. By repeated absorption-emission cycles photons thermalize to the temperature of the dye solution, and in a single microsite we observe a photon Bose-Einstein microcondensate. Effective interactions between the otherwise nearly non-interacting photons are observed due to thermo-optic effects, and in a double-well system tunnel coupling between sites is demonstrated, as well as the hybridization of eigenstates. Prospects of the new experimental platform include photonic structures in which photons thermalize into entangled manybody states[5].**

Periodic potentials for light are at the core of proposals for Mott insulator physics for light, topological effects, as well as driven-dissipative phase transitions[13-18]. Exciton-polariton experiments, involving mixed states of matter and light under conditions of strong coupling, have used permanent semiconductor micro-structuring, as molecular beam epitaxy, metal depositing techniques, and mirror patterning[19,20], to demonstrate double-well and periodic potentials[6-9]. In the regime of weak light-matter coupling, thermalization and Bose-Einstein condensation of a photon gas has been achieved in a high finesse microcavity containing dye molecules in liquid solution[11,12].

Here we demonstrate a microstructuring technique that allows to generate variable potentials for light within an optical high-finesse microcavity. The long photon lifetime enables the thermalization of photons and the demonstration of a microscopic photon condensate in a single localized site. We observe effective photon interactions as well as tunnel coupling between two microsites. The associated hybridization of eigenstates of the double well system is monitored spectroscopically.

The scheme for thermo-optic imprinting of potentials is shown in Fig.1a. Within a microcavity of finesse near 35000, variations of the refractive index are induced through irradiation with a laser beam inducing heat from absorption in a 30nm thick silicon layer below one of the mirror surfaces. A thermosensitive polymer (PNIPAM)[21], which undergoes a reversible phase transition to a phase with higher refractive index[22] when heated above 305K within a narrow temperature range of 0.2K, is added to the dye solution between the mirrors. Local heating correspondingly increases the optical length between the mirror surfaces, which is equivalent to a local potential drop for the photon gas in the paraxial limit[11]. This can be understood from the larger optical wavelength, corresponding to a smaller photon energy, required to locally match the mirrors' boundary conditions. Transverse temperature patterns within the dye polymer film are created by scanning a focused laser beam with a two-dimensional galvo scanner over the absorbing silicon coating, while an acousto-optic modulator allows for intensity control, making the technique useful for the creation of variable potentials for light.

Thermalization of the photon gas confined in the microcavity is achieved similarly as described previously[11]. Briefly, the short mirror spacing effectively leads to a low-frequency cutoff for the photon gas. By repeated absorption and re-emission processes on the dye molecules, photons thermalize to the rovibrational temperature of the dye, which is near room temperature. As the spacing between longitudinal modes is of order of the emission width of the dye (rhodamine in water solution), the thermalization process leaves the longitudinal mode number constant. Correspondingly, the photon gas is effectively two-dimensional, with only transverse modal quantum numbers varied.

The dye solution is pumped with a beam near 532nm wavelength, chopped to 1μs pulses with a 50Hz repetition rate, to reduce the effects of pumping into dye triplet states. The pulse length is more than two orders of magnitude above the thermalization time[26]. As an example for the achieved spatial structuring of the confined photon gas in the thermo-optically imprinted potential, Fig.1b shows the cavity emission for periodic patterns (left and middle) and a non-periodic pattern (right).

First, we have investigated thermalization of the photon gas in a single microtrap. The top panel of Fig.2a shows typical spectra of the emission of such a microsite for a low-frequency cutoff wavelength of 595nm and different photon numbers in the cavity. The depth of the potential is h·7.5THz, corresponding to $1.21 \cdot k_B T$ at T= 305K, and the observed mode spacing is $\Omega/2\pi \cong 1.18$THz near the trap bottom, which reduces for the higher transverse modes due to deviations from a two-dimensional harmonic oscillator potential. In the thermalized case, we expect the mode spectrum to be Bose-Einstein distributed following

$$n(u_i) = \frac{g(u_i)}{\exp\left(\frac{u_i - \mu}{k_B T}\right) - 1},$$

where $u_i$ (with $u_i = i \cdot \hbar\Omega$ for a harmonic potential) denotes the excitation energy of the i-th trap level with respect to the low-frequency cutoff, μ the chemical potential, and $g(u_i)=2(i+1)$ the degeneracy. The bottom panel of Fig.2a gives spectrally binned data, yielding the trap levels' population, which well agrees with the predictions within experimental uncertainties. We observe a spectrally broad thermal cloud above the low-frequency cutoff, upon which above a critical photon number a Bose-Einstein condensed peak at the position of the cutoff emerges. The observed critical photon number of 67.8(1.6) is three orders of magnitude lower than in previous work with "macroscopic" photon traps[11], and within uncertainties agrees with the expected value $N_c$=67.5 (Methods). The observation of thermalized spectra shows that in photon microtraps loss through e.g. mirror transmission can be kept sufficiently low that within the cavity lifetime photons relax to a near equilibrium distribution. The observed condensate diameter of ~3μm is close to the harmonic oscillator ground mode size, and we typically use a 15μm pump beam diameter. For the future, such photon microcondensates are expected to be attactive systems to study few-particle physics effects[23].

To investigate the effect of thermo-optic interactions induced by both the pump beam and condensate photons[11,24,25], as understood e.g. from residual non-radiative decay channels of excited dye molecules, we have temporally resolved the microcavity emission frequency during a pump pulse, see Fig.2b. We observe an increasing blue shift of the condensate frequency, corresponding to a decreasing refractive index of the dye-polymer solution. As the temperature increases during the pump pulse, we conclude that the effective thermo-optical coefficient *dn/dT* is negative. The polymer in steady state has a positive thermo-optic coefficient, but due to the slow (~500ms) timescale of its response on the 1μs short time scale of a pump pulse the thermo-optic properties of the water solvent (with *dn/dT*<0) dominate. We additionally have characterized the self-interaction of the condensate by measuring the variation of the mode diameter during one pulse, see Fig. 2c. We observe a diameter increase, and for sufficiently large pump laser spots, this variation becomes independent on details of the pump geometry, and therefore can be attributed to condensate self-interactions. Using a Gross-Pitaevskii model for the retarded thermooptic effect we derive a self-interaction parameter of $g_{eff}$=2.5(8)x$10^{-5}$ accumulated during 1μs from our data (Methods), which is even smaller than previously reported for a rhodamine-methanol medium[11].

To study tunneling we employ double-well potentials with a typical spacing between microsites of 8-15 μm. The potentials are prepared with a well depth so shallow that only a single mode is trapped, of frequencies $\omega_1$, $\omega_2$ for the two microsites 1,2 respectively. One of the sites, say site 1, is pumped, and we slightly red-shift the initial emitter frequency of this pumped site with respect to that of site 2, so that during the pump pulse the thermo-optic effect induced by the pump beam will tune the sites into resonance, as indicated in Fig.3a. Corresponding data for the time-resolved fluorescence emitted from the individual microsites is shown in Fig.3b. As the sites are tuned into resonance, we observe tunneling between microsites, with the number of tunneled photons again decreasing when the sides shift out of resonance. The resonance width increases for a reduced spacing between sites, as understood from the then larger tunnel coupling. The coupling strength derived from the width of the resonance (Methods) shown in Fig.3c follows the expected exponential scaling with the distance between microsites. At the smallest investigated distance near 8 microns, the tunnel coupling reaches a value of 16.7(1.1) GHz. The influence of the coupling on the emission frequency was explored by time-resolved spectroscopy of the emission from site 1 during the frequency chirp caused by the thermo-optic effect. Figure 4a was recorded for a tunnel coupling between sites of $J/2\pi$ =1.7GHz, where no coupling-induced splitting of eigenstates is resolved, while the measurement shown in Fig.4b gives data for $J/2\pi$ =16.7GHz, where a clear splitting is visible. The emission frequencies well follow expectations for an avoided crossing with the corresponding coupling (dashed lines), as readily obtained when modeling the double-well system with a set of coupled Schrödinger equations[27]:

$$i\hbar\dot{\psi}_1(t) = \hbar\omega_1 \cdot \psi_1 + \hbar J \cdot \psi_2,$$

$$i\hbar\dot{\psi}_2(t) = \hbar J \cdot \psi_1 + \hbar\omega_2 \cdot \psi_2,$$

where $\psi_{1,2}$ denote optical wavefunctions in the corresponding sites. The eigenstates of the coupled system are $\psi_+ = (\sin(\theta)\psi_1 + \cos(\theta)\psi_2)$, $\psi_- = (\cos(\theta)\psi_1 - \sin(\theta)\psi_2)$, with eigenenergies $E_\pm = \hbar(\omega_1 + \omega_2)/2 \pm \sqrt{(\hbar\Delta\omega/2)^2 + J^2}$, where $\theta = \arctan(-2J/\Delta\omega)/2$ denotes the mixing angle and $\Delta\omega = \omega_1 - \omega_2$. Figure 4c gives the expected emission signal $\propto |\langle\psi_\pm|\psi_1\rangle|^2$ from site 1 for $J/2\pi = 16.7$GHz versus the detuning. The experimental data of Fig.4b indicates an even further suppression of the emission on the weaker side of the two branches where the eigenstates mostly overlap with the unpumped site, an issue attributed to photon losses.

The right hand side of Fig.4b gives camera images of the emission of the double-well system for mode frequencies of site 1 below (top) and above (bottom) that of site 2. For the former case, (i), we observe a non-vanishing emission in the region between sites (in the classically "forbidden" region), indicating that the system here is in the symmetric eigenstate $\psi_+$, while in the latter case (ii) no emission is observed in that region, as expected for photons in the antisymmetric state $\psi_-$. When spatially overlapping the emission of the two microsites, the observed fringe pattern shifts by π in phase when tuning from below to above the tunneling resonance (from case (i) to (ii), respectively). Figure 4b (bottom right), shows the different phase factors of symmetric and antisymmetric eigenstates respectively of the "photonic molecule", which are well in agreement with expectations.

To conclude, we have demonstrated variable potentials for light in an ultrahigh-finesse optical cavity generated using thermo-optic imprinting, a technique compatible with thermal equilibration. A microscopic photon Bose-Einstein condensate with a critical photon number of 68 was observed in a single microtrap. Effective photon interactions as well as photon tunneling and hybridization of eigenstates in a double-well potential has been achieved.

For the future, we expect novel applications of thermo-optic patterning within low-loss resonators also in other fields of photonics. Prospects of photon condensates in lattice potentials include the population of entangled states[5,28]. Other than in ultracold gas systems[1], loading and cooling in our system proceeds throughout the lattice manipulation time, and when constituting the system ground state quantum manybody states can be populated in a thermal equilibrium process.

†Present address: Institute for Quantum Electronics, ETH Zürich, Auguste-Piccard-Hof 1, 8093 Zurich, Switzerland

**Figures:**

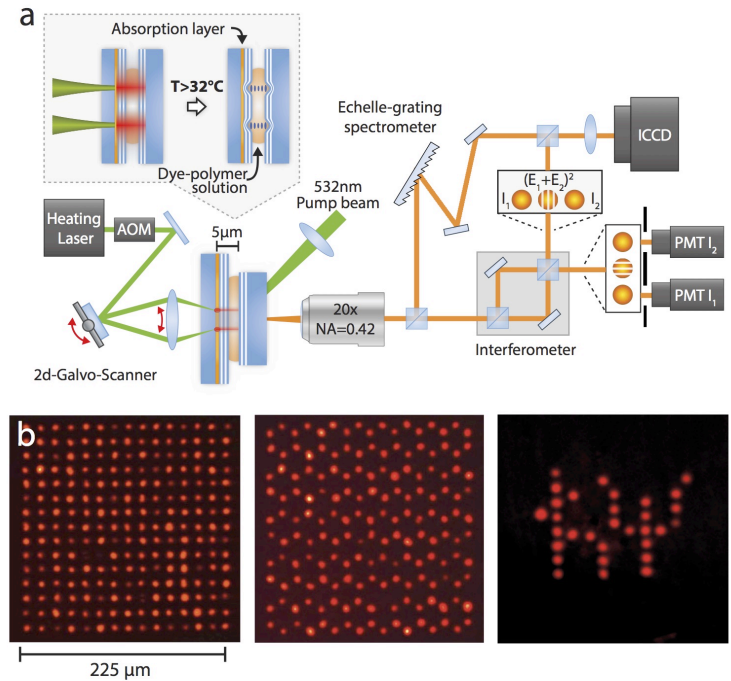

**Figure 1 | Setup and lattice realizations. a,** Experimental approach to realize variable potentials for photons within a high-finesse microcavity. Optical radiation is absorbed in a thin silicon layer placed below one of the cavity mirror's dielectric coatings. The induced heat locally increases the index of refraction of the dye-polymer solution within the microcavity, and here enhances the optical length. This results in an effective local attractive potential for the photon gas. Complex trapping potentials are realized by scanning an external "heating" laser beam to imprint the desired transverse temperature modulation onto the absorbing silicon layer. The microcavity is pumped from the reverse side, and its emission can be analyzed spatially, interferometrically, and spectrally. **b,** Experimentally observed images of the microcavity emission for the photon gas in lattices with rectangular (left) and hexagonal (middle) geometry respectively, and one non-periodic pattern showing the letters "$\hbar\psi$" (right).

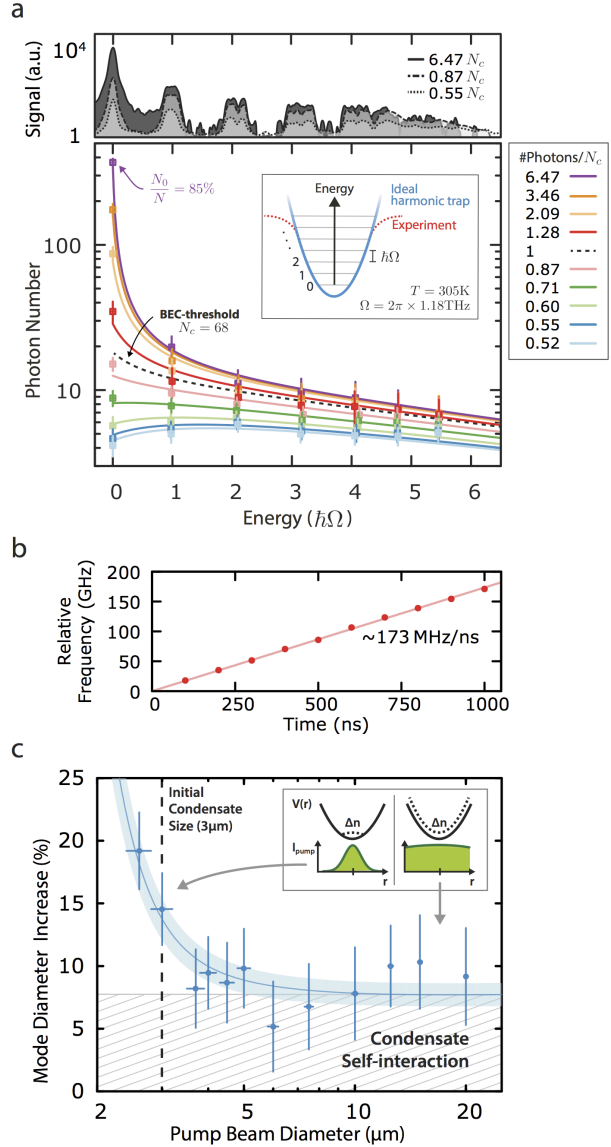

**Figure 2 | Thermalization, condensation and photon self-interactions in a single microsite. a,** Spectral intensity distribution of the microcavity emission for different photon numbers in units of the critical photon number $N_c$, with the top panel showing spectrometer signals and the bottom panel spectrally binned signals (dots) as to determine the trap level populations, together with theoretical expectations (lines). We observe a spectrally sharp condensate peak at the position of the cavity cutoff on top of a broad thermal cloud distributed over all bound 7 trap levels. For the higher trap levels the level spacing reduces due to deviations from harmonic trapping (see also inset). **b,** Relative emission frequency of the dye microcavity versus time, showing a linear chirp attributed to thermal lensing induced predominantly from the pump beam. The 1μs duration of the used pump pulses is below the thermal time constant so that no saturation of the chirp is visible (Methods). This data was recorded for a single microsite, but with the same pump geometry as used for the measurements shown in Figs.3 and 4. **c,** Observed condensate mode size within a pump pulse versus diameter of the pump beam. The photon number in the condensate mode is kept constant for all measurements by adjustment of the pump beam power. The broadening observed for small pump spot diameters of size comparable to the ground mode is attributed to heating directly from the pump beam. For larger diameters the variation of the optical cloud diameter becomes independent from the pump diameter, as well understood from thermo-optic effects due to condensate photons. The latter effect is due to heating in course of the absorption re-emission cycles of the dye in the presence of the finite quantum efficiency. A Gross-Pitaveskii equation model is used to describe the findings (Methods).

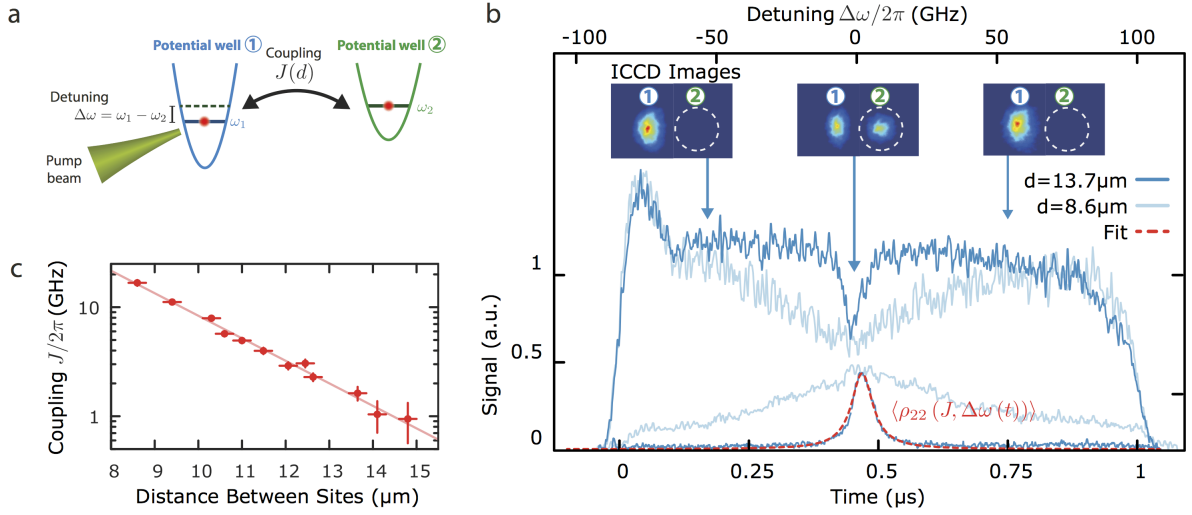

**Figure 3 | Tunneling between two microsites. a,** Schematic of the coupling and **b**, time dependence of the emission observed from microsite 1 (top) and site 2 (bottom) for a spacing of 13.7(2) μm (dark blue lines) and 8.6(2) μm (light blue lines) respectively. As the mode energy of the pumped site 1 increases due to thermo-optic effects, the two trap levels align in energy and photons tunnel between sites. Note that the 230 MHz bandwidth of the used photomultiplier detector is not sufficient to resolve temporal oscillations between sites. The top scale gives the corresponding detuning between sites assuming a linear chirp of the mode frequency of site 1 with time, see the measurement shown in Fig.2b. The insets give camera images recorded at corresponding times. **c,** Extracted tunnel coupling versus distance between microsites.

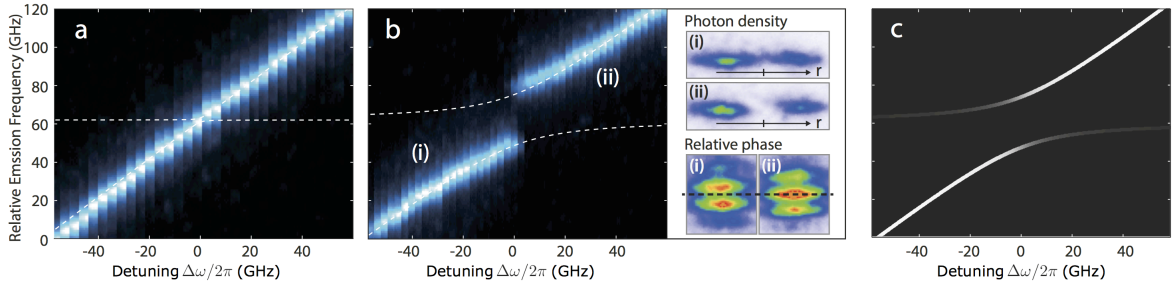

**Figure 4 | Eigenstate hybridization in double-well.** Spectrum of microcavity emission of site 1 versus the detuning between the sites for a tunnel coupling of $J/2\pi =1.7$ GHz (**a**) and $J/2\pi =16.7$ GHz (**b**). The experiment was realized with a spectrometer operating in a gated mode with a 20ns long detection period using repeated scans with a variable position of the gate. The dashed lines show the expected location of dressed eigenstates for the corresponding coupling. The two top plots on the right side of (**b**) give camera images for the emission in both sites for a detuning of site 1 (i) below and (ii) above that of site 2, and the bottom plots corresponding interferometer signals. **c**, Theory prediction for the emission from site 1 for a $J/2\pi =16.7$ GHz tunnel coupling between sites in the absence of losses .

**Methods:**

**Experimental setup and procedure.**

The used microcavity consists of two plane highly reflective mirrors (specified reflectivity above 99.991% after reflection of the two cavity mirrors in the spectral range of 520-590nm wavelength) spaced by a distance $D_0$=1.98μm. We use a plane-mirror microcavity, so that effective trapping potentials for the photon gas are solely due to thermo-optic transverse index variations. The microcavity is filled with a solution of poly N-isopropylacrylamide (PNIPAM), (c=8%), a thermosensitive polymer, and rhodamine 6G dye (c=0.04%, corresponding to $10^{-3}$ mol/l) solved in water and Ammonyx-LO (c=5%) as a surfactant. Water was used as a solvent since the polymer here solves well, despite the somewhat smaller quantum efficiency of the rhodamine dye in this environment (≈85%-95%) in comparison to that obtained with the ethylene glycol solvent used in our earlier experiments[11]. Above 32.4 °C temperature, the polymer within a temperature range of ≈0.2°C undergoes a phase transition to a state with collapsed polymer chains, and the associated molecular transport increases the specified refractive index from n≈1.35 to n≈1.46[29]. We have not seen evidence for a reduction of the cavity finesse caused by the dye-polymer phase transition. The timescale of the refractive index change within the microcavity environment was experimentally determined by monitoring the induced variation of the cavity emission frequency upon local heating to ≈500ms. Between glass substrate and dielectric coating of one of the cavity mirrors a 30 nm thick silicon coating is placed, which at 532nm wavelength absorbs 30% of the incident irradiation to convert the pattern of a transversally spatially modulated optical pattern into a temperature profile. A corresponding pattern is induced by a beam near 532 nm wavelength of 5-10 mW power focused to 2 μm diameter that is scanned transversally with a 2D galvo scanner. This scan is done on a 100ms timescale, which is fast compared to the above mentioned timescale of the polymer phase transition. The beam power is controlled with an acousto-optic modulator. The local variation of the optical length between mirror surfaces induced by the refractive index change leads to a transversally varying potential in the microcavity. Due to the high reflectivity of the dielectric coating, transmission of the auxiliary "heating" beam into the microcavity is suppressed by four orders of magnitude, allowing for a control of the cavity photon number independently from the "heating" laser beam. Photons are injected into the microcavity by pumping with a further laser beam near 532nm wavelength from the reverse cavity side at typically 45° angle to the optical axis. The typical used pump power is up to 2 W on a 15μm beam diameter for the single well measurements shown in Fig.2a, while to pump one site of the double well measurements shown in Figs.3 and 4, as well for the measurement shown in Fig.2b, an elliptical beam of diameters 3 and 20 μm along the short and long axis respectively was used.

Photons trapped in the microcavity acquire a thermalized distribution from absorption and re-emission processes on dye molecules, provided that the cavity lifetime exceeds the thermalization time[26,30]. Rapid decoherence from frequent collisions of dye molecules with the solvent here prevents a coupling of the phases of dipole and photon[11], so that we can assume to be in the weak coupling regime of photons and molecules. The small spacing, corresponding to 4.5 optical wavelengths in the medium, causes a large frequency spacing between longitudinal cavity modes, which is comparable with the emission width of the dye molecules. We observe that to good approximation only photons of a fixed longitudinal mode, q=9 here, populate the cavity, making the system effectively two-dimensional. This also introduces an effective low-frequency cutoff near $\hbar\omega_{cutoff} \cong 2.1$ eV, with $\omega_{cutoff} = 2\pi c/\lambda_{cutoff}$, $\lambda_{cutoff} \cong 595$ nm. The pump beam is acousto-optically chopped into 1μs long pulses at a typically 50 Hz repetition rate to suppress pumping into dye triplet states. The pulse duration exceeds the thermalization time, which is of order of the 3ns upper electronic state natural lifetime of the rhodamine dye molecule, or far above the threshold in the ps regime[26], by at least two orders of magnitude, i.e. from the point of thermalization of the photon gas the experiment operates in a quasi-cw mode. The obtained maximum condensate fractions were near 85%, corresponding to the top spectrum shown in the upper and lower panels of Fig.2a, while at higher photon numbers deviations from equilibrium spectra increase[31].

With the thermo-optic method in the dye polymer microcavity potential well depths up to h·9.5 THz, corresponding to ~1.5 $k_B T$, have been achieved, and the trap frequencies Ω/2π of individual microtraps are in the range 0.2 to 2 THz. We have realized trap sizes down to 3μm diameter, a value attributed to be limited by transverse heat diffusion due to the nonvanishing distance between the silicon heating layer and the dye-polymer solution determined by the thickness of the dielectric mirror layers. The quoted value for the maximum well depth corresponds to a refractive index change near Δn≅0.07 (i.e. below the expected maximum possible difference of Δn≅0.11, thus reducing effects of saturation) of the thermo-sensitive polymer at the phase transition. We here have assumed that of the optical distance between the reflecting mirror surfaces of q=9 half waves a fraction $q_0$=6.7 is within the dielectric coatings, so that the actual optical thickness of the dye polymer solution film is q-$q_0$=2.3 optical half waves, with a refractive index n≅1.35 below the phase transition.

Each data point of the emission spectrum of a single microsite shown in Fig. 2a (top) is the average of 2000 individual measurement results. The resolution of the used spectrometer is 30 GHz. The observed spacing between the lowest transverse modes is nearly equidistant, so the generated potential near the trap bottom can well be assumed to be harmonic. This spacing of Ω/2π ≅ 1.18 THz, corresponding to the trap vibrational frequency, translates to an effective radius of curvature of the reflecting surfaces of $R_{eff} = 2c^2/(n^2 D_0 \Omega^2) \cong 0.9$ mm. One can show that the photon gas in a spherically curved mirror resonator is formally equivalent to a two-dimensional gas of harmonically confined, massive particles with an effective mass $m_{ph} = \hbar\omega_{cutoff}/(c/n)^2$, for which it is known that a Bose-Einstein condensate exists at thermal equilibrium conditions[11]. The expected critical particle number is

$$N_c = \sum_{u=u_1,u_2,\ldots,u_6} \frac{2(u/\hbar\Omega+1)}{e^{(u/k_B T)}-1}, \qquad (1)$$

which for the above quoted value for Ω and the ambient temperature T=305K yields $N_c$≅67.5. This value is three orders of magnitude below the value reported in our earlier "macroscopic" photon Bose-Einstein condensation experiment, carried out with two resonator mirrors of 1m spherical curvature[11]. We note that the here reported experiment operates far from the thermodynamic limit, as there only are 7 trapped energy levels (56 trapped levels in total including also the polarization degeneracy). Despite this small number of energy levels, our experimental observations are in good agreement with Bose-Einstein condensation theory. A visible small overpopulation of the higher order modes for the spectra with low photon numbers is attributed to imperfect spatial filtering of the fluorescent emission from the microsite, causing some stray light from continuum modes in the presence of a finite spectral resolving power to reach the detector.

The measurements shown in Fig.2b were recorded by spectrally analyzing the emission of a single microsite in a time-resolved way. For this, the ICCD gate time was varied between 100ns and 1μs, and the corresponding variation of the observed spectral width was monitored with an Echelle grating spectrometer of 9 GHz spectral resolution. We observe a to good accuracy linear frequency chirp to higher eigenfrequencies which within the pulse time does not approach a steady state. This data was recorded for the same pump beam geometry as used for the measurements shown in Figs. 3 and 4, and is attributed to be dominated by thermo-optic effects induced from heating from the pump beam pulse. In the data of Fig.2b we observe a linear slope with a chirp rate of $d\nu/dt$ =173(4) MHz/ns. This allows us to calibrate the time axis shown in the double-well data of Fig. 3b to a frequency axis giving the detuning between sites, as shown on the top of the diagram. The quoted value for the chirp rate can be used to estimate a local heating rate $dT/dt$ of the dye-polymer solution during the pump pulse. We here arrive at $dT/dt=(-1/\kappa)\cdot n/\nu \cdot d\nu/dt \approx 4.6 K/\mu s$, where $\kappa$ denotes the thermo-optic coefficient of water ($\kappa \approx -1\cdot 10^{-4}$ 1/K).

As described in the main text, the refractive index change due to the polymer phase transition is much larger and of different sign than refractive index variations from the thermo-optical effect in water. The typical timescales of the two effects is 0.5s for refractive index change due to the polymer phase transition and $\tau \approx 4\mu s$ for the usual thermo-optic effect at a few microns optical beam diameter differs by an even larger factor, so that the transient response is dominated by the latter effect. Note that the timescale of the thermooptic effect is dependent on the geometry of the system, and increases for larger optical beam sizes. Fig. 2b shows the corresponding experimentally observed increase of the condensate frequency within the 1 μs pulse. These above conclusions are furthermore supported by the observation that within the 1μs long pump pulse we for all pump beam diameters (i.e. both in the case of the observed thermo-optic effect being dominated by heating from the pump pulse itself and the cavity photons) observe an increase of the ground mode diameter, as expected for *dn/dT*<0.

The measurements shown in Fig.2c were conducted to characterize thermo-optic effects in the dye-polymer cavity environment in more detail. For those measurements potentials with reduced depth were used, so that only a single level was trapped (as for the data shown in Figs. 3 and 4). In the experiment the diameter of the macroscopically occupied mode is measured alternating at the beginning and the end of the 1μs long pump pulse, with the gate of the ICCD camera set to 50ns. The vertical scale gives the observed relative broadening of the emitted condensate mode within a pump pulse. Each data point gives the average of the results of 200 individual measurements. The described procedure is carried out for different pump beam diameters, while keeping the photon number in the macroscopically occupied mode fixed ($N_0$=50000±4000). For that the pump power was increased quadratically with the pump beam diameter. We note that the pump power and correspondingly the photon number in the condensate mode for this measurement was so high that no photon clouds at thermal equilibrium conditions could be observed. Due to the presence of only a single trapped level, thermalization was not relevant for this measurement.

To model the system, we start with a two-dimensional Gross-Pitaevskii equation

$$\left(-\frac{\hbar^2 \nabla^2}{2m_{ph}} + V(\vec{r};t) + E_{int}(\vec{r};t)\right)\psi(\vec{r};t) = \mu\psi(\vec{r};t), \qquad (2a)$$

where $V(\vec{r},t)$ is an effective trapping potential for photons and μ the chemical potential. Thermo-optic effects are assumed to be slow with respect to the timescale on which the photon wavefunction reaches a steady state. Correspondingly, eq.2a has the form of the stationary Gross-Pitaevskii equation with a residual time-dependence due to the slow variation of the thermo-optical self-interaction. The interaction energy due to the refractive index change of the solution from heating through condensate photons can be written as

$$E_{int} \cong -m_{ph}(c/n_0)^2 \frac{\Delta n}{n_0}, \qquad (2b)$$

where $n_0$ is the refractive index of the solution at temperature $T_0$ of the mirrors, which are regarded to act as a heat sink. Further, $\Delta n = \left(\frac{\partial n}{\partial T}\right) \cdot (T(\vec{r};t) - T_0) \equiv A \cdot (T(\vec{r},t) - T_0)$, where $\Delta n(\vec{r},t)$ and $T(\vec{r},t)$ denote the refractive index variation (with $\Delta n << n_0$) and temperature respectively at the corresponding transverse position averaged over the cavity length, and $(\partial n/\partial T)$ is the thermo-optic coefficient of the solution. As we do not observe evidence for an (ultrafast) Kerr interaction[11,24,25] at the present level of accuracy with the here used solution, no corresponding term is added to eq.2b. In our model, we assume that the temperature profile of the cavity relaxes towards a steady state that is determined by the intensity of the light field on a timescale τ. The corresponding time-evolution of the temperature is described by

$$\frac{dT(\vec{r},t)}{dt} = -\frac{1}{\tau}\left(T(\vec{r},t) - T_0\right) + B|\psi(\vec{r},t)|^2, \qquad (2c)$$

and has the formal solution[32]

$$T(\vec{r},t) = T_0 + B\int_{-\infty}^{t} N_0 |\psi(\vec{r},t')|^2 e^{-(t-t')/\tau} dt'. \qquad (2d)$$

With the initial condition $|\psi(\vec{r},t)|^2 \equiv 0$ for t<0, we arrive at a Gross-Pitaevskii equation for a temporally retarded interaction term

$$\left(-\frac{\hbar^2 \nabla^2}{2m_{ph}} + V(\vec{r};t) + \frac{\hbar^2}{m_{ph}}\frac{g_{stat}}{\tau} N_0 \int_0^t |\psi(\vec{r},t')|^2 e^{\frac{t'-t}{\tau}} dt'\right)\psi(\vec{r};t) = \mu\psi(\vec{r};t), \qquad (2e)$$

where $g_{stat} = -\tau ABm/\hbar^2$ denotes the dimensionless two-dimensional interaction constant expected at stationary conditions due to thermo-optic self-interactions.

Refractive index changes induced by heating from the pump beam can be accounted for in an analogous way, by first introducing a heating term into eq. 2c that has the spatial transverse profile of the Gaussian shaped pump beam. We arrive at an extra potential energy $V_{pump}(\vec{r};t) = \xi(t) \cdot I_{0,pump} \exp(-2r^2/w_{pump}^2)$, where $I_{0,pump} = 2P_{pump}/\pi w_{pump}^2$, with $w_{pump}$ denoting the radius and $P_{pump}$ the power of the pump beam, and $\xi$ is a proportionality factor that for $t \ll \tau$ (similarly as the third term in eq. 2e) scales linearly with the length of the pump pulse. When including a harmonically shaped photon trapping potential $V_{harmon}(\vec{r}) = (m_{ph}/2)\Omega^2 r^2$, the total trapping potential can be written in the form

$$V(\vec{r};t) = V_{trap}(\vec{r}) + V_{pump}(\vec{r};t) \cong \frac{m_{ph}}{2}\Omega'(t)^2 r^2 + \xi(t) \cdot I_{0,pump}, \tag{2f}$$

with $\Omega'(t) = \Omega\left(1 - 2\xi(t) \cdot I_{0,pump}/\Omega^2 m_{ph} w_{pump}^2\right)$ for $\Omega'/\Omega \ll 1$.

We next restrict ourselves to the case of times t much below the retardation time constant $\tau$ and sufficiently small interactions. In this limit, only small modifications of the wavefunction due to interactions are expected, so that the approximation $|\psi(\vec{r};t')|^2 \cong |\psi(\vec{r};0)|^2$ in the interaction term is justified. Correspondingly, this term simplifies to $E_{int}(\vec{r};t) = \frac{\hbar^2}{m_{ph}} g_{eff}(t) N_0 |\psi(\vec{r};0)|^2$ with the time-dependent effective interaction constant $g_{eff}(t) = \frac{t}{\tau} g_{stat}$. In this limit, the problem can for the harmonically trapped case be solved analytically with a variational method, as described in Ref. 33. We begin by writing the energy of the system in the functional form

$$E(\psi) = \int d\vec{r} \left( -\frac{\hbar^2}{2m_{ph}} |\nabla \psi(\vec{r};t)|^2 + V(\vec{r};t)|\psi(\vec{r};t)|^2 + \frac{1}{2}\frac{\hbar^2}{m_{ph}} g_{eff}(t) N_0^2 |\psi(\vec{r};t)|^4 \right). \tag{2g}$$

Note that minimization of $(E-\mu N_0)$ with respect to $\psi^*$ at a fixed value of the chemical potential $\mu$ gives the Gross-Pitaevskii equation of eq. 2e. To find the condensate wavefunction, we use a Gaussian trial function of width $w = w_{0,cond} + \Delta w_{cond}$, where $w_{0,cond}$ denotes the width of the harmonic oscillator wavefunction (as obtained with no interactions), and minimize the energy in the presence of interactions with respect to variations in $w$. When also accounting for the above described thermo-optic effects induced by the pump beam, we find a relative broadening of the condensate diameter of

$$\frac{\Delta w_{cond}}{w_{0,cond}} \cong \frac{\xi \cdot P_{pump}}{\Omega^2 m_{ph}^2 \pi w_{pump}^4} + \frac{N_0}{8\pi} \cdot g_{eff}(t). \tag{2h}$$

where we have assumed that $\Delta w_{cond} \ll w_{cond}$. The first term is due to thermo-optic effects induced directly by heating from the pump beam, which rapidly decays with increased size of the pump beam diameter. The contribution from photon self-interactions can then readily be identified from the asymptotic value at large pump beam diameters, which is described by the second term (as indictated by the hatched area in Fig. 2c). In contrast to the case of the Thomas Fermi regime (interaction energy dominating over the kinetic energy) for which in 2D systems the condensate diameter scales with $(N_0)^{1/4}$, the here presented system is in the opposite regime of the kinetic energy dominating over the kinetic energy, for which one finds a linear increase of the condensate diameter the particle number $N_0$.

At the used t=1μs thermooptic interaction time, corresponding to the length of a pump pulse, the observed asymptotic value of the condensate mode broadening at large pump beam diameter reaches 8(1)%. Using eq.2h we derive an accumulated effective interaction constant $g_{eff} = 4(2) \times 10^{-5}$. In addition to the described analytic estimation, we have also performed a full numeric evaluation of the Gross-Pitaevskii equation, from which we find a numeric value of the interaction constant of $g_{eff} = 2.5(8) \times 10^{-5}$ for the observed condensate broadening. The latter value corresponds to a static interaction constant $g_{stat} = \frac{\tau}{t} \cdot g_{eff} = 1.0(3) \times 10^{-4}$, when using the above quoted value of a τ=4μs thermal time constant. These quoted values for the dimensionless interaction constant are below that reported in earlier experiments of our group for a different experimental situation [11], as understood mainly (i) from the use of water as a solvent in the present experiment with its comparatively low thermooptic coefficient (ii) the cutoff being at a relatively long wavelength resulting in a small photon absorption rate and (iii) the here quoted value referring to the influence of the thermooptic interaction during a single pump pulse.

The tunneling couplings J between potential wells were determined from the spectra shown in Fig.3b using a Bloch equation model accounting for a pump rate λ of site 1 and a loss rate Γ:

$$\dot{\rho}_{11} = \lambda - \Gamma \rho_{11} - 2J \operatorname{Im}(\rho_{12}) \;, \tag{3a}$$

$$\dot{\rho}_{22} = -\Gamma \rho_{22} + 2J \operatorname{Im}(\rho_{12}) \;, \tag{3b}$$

$$\dot{\rho}_{12} = -(\Delta\omega + \Gamma)\rho_{12} - iJ \operatorname{Im}(\rho_{22} - \rho_{11}) \;, \tag{3c}$$

where $\rho_{11}$, $\rho_{22}$ denote the populations in sites 1,2 respectively, $\rho_{12}$ (=$\rho_{21}^*$) the corresponding off-diagonal matrix element, and $\Delta\omega = \omega_1 - \omega_2$. One readily finds the stationary analytic solutions

$$\rho_{11}^{stat} = \frac{\lambda}{\Gamma}\left(1 - \frac{2J^2}{\Delta\omega^2 + \Gamma'^2}\right) \;, \tag{4a}$$

$$\rho_{22}^{stat} = \frac{\lambda}{\Gamma} \cdot \frac{2J^2}{\Delta\omega^2 + \Gamma'^2} \;, \tag{4b}$$

$$\rho_{12}^{stat} = \frac{-i\Gamma - \Delta\omega}{\Delta\omega^2 + \Gamma^2} \cdot 2\left(\rho_{22}^{stat} - \rho_{11}^{stat}\right) \;, \tag{4c}$$

with the saturation broadened resonance width $\Gamma' = \sqrt{\Gamma^2 + 4J^2}$.

From the observed value for the resonant tunneling ratio for the spectra as shown in Fig.3b we determine the ratio *J/Γ* of tunnel coupling and loss rate using the expected ratio $\rho_{11}(\Delta\omega = 0)/\rho_{11}(\Delta\omega \to \infty) = 1 - 1/\left[2(1+(\Gamma/2J)^2)\right]$ for each spectrum of a certain distance between microsites. This allows us to in the next step from the observed width of the spectra, as obtained from a fit, determine the corresponding value for the tunnel coupling *J*, see Fig.3c for corresponding results. This readily also allows us to determine the loss rate, and after averaging over the results of the different spectra a value of Γ=11(5) GHz is obtained.

The spectral data shown in Figs.4a and 4b was obtained by using a gated ICCD camera detector (20ns gate time) monitoring the emission after the Echelle monochromator to analyze the microcavity emission. Such measurements were repeated for different delay times after the onset of the pump pulse to allow for the analysis of the temporal variation of the microcavity emission during the frequency chirp of the relative detuning of microsite 1 relatively to site 2 from the thermo-optic effect. The expected value for the population of dressed eigenstates of the double-well system shown in Fig.4c was derived from projecting the eigenstates of the system onto the left well (site 1). This method gives the expected value for the relative values of the emission of the two sites in the absence of losses.

**Acknowledgements**

We acknowledge financial supported by the DFG (CRC 185) and the ERC (INPEC).